# Managing Knowledge to Enhance Learning

## Philippe A. Martin*

Eurecom (France) and Griffith University (Australia),
Griffith Uni. - School of ICT - PMB 50 GCMC -  QLD 9726 Australia
E-mail: kme @ phmartin.info.

*Corresponding author

**Abstract:** The article first summarizes reasons why current approaches supporting Open Learning and Distance Education need to be complemented by tools permitting lecturers, researchers and students to cooperatively organize the semantic content of Learning related materials (courses, discussions, etc.) into a fine-grained shared semantic network. This first part of the article also quickly describes the approach adopted to permit such a collaborative work. Then, examples of such semantic networks are presented. Finally, an evaluation of the approach by students is provided and analyzed.

**Biographical notes**: After his Ph.D. at the INRIA (France) and its postdoc at the University of Adelaide (Australia),  Philippe A. Martin  worked as (senior) researcher/lecturer at Griffith University and the DSTC (Australian W3C office). Since 2008 Dr Martin is project leader at Eurecom (France). He mainly worked in knowledge representation and management.

## 1.  Introduction

Most Semantic Learning projects (Stutt & Motta, 2004; Devedzic, 2004) and most Learning Object related standards or practices (Downes, 2001; IEEE LTSC, 2001; Hodgins, 2006; Tane et al., 2003) rely on a rather *coarse grained indexation* of informal data (natural language sentences, images, etc.) by simple meta-data. Conceptual categories or even mere keywords are manually or automatically associated to relatively big chunks of informal data (typically a whole document, and almost always more than one sentence). In *fine-grained approaches*, the data (learning materials or very-detailed personalized user models) are *represented and organized into a formal or semi-formal semantic network without redundancies*. To do so, the content of the source materials is decomposed into a set of data units which ideally are all irreducible units (i.e., conceptual categories or formal/informal stand-alone statements) and inter-related by semantic relations (e.g., relations of specialization, argumentation, instrumentation, correction, authorship, spatial/temporal location and modality). Some of these networks are fully formal, very difficult to create, and difficult to read or search directly. This is typically the case for  tutor systems, for example for Halo (Friedland et al., 2004) which permits to solve some chemistry test questions automatically. Most other semantic networks, especially those of projects using Concept Maps (Novak, 1984; Novak & Gowin, 2004) or Topic Maps (Scott & Johnson, 2005), are mostly informal and difficult to re-use for information retrieval and comparison.

In (Martin & Eboueya, 2008) I showed that none of the current main knowledge sharing/retrieval approaches permit lecturers, researchers and students to collaboratively build a "fine-grained semi-formal normalized source/creator-keeping readable semantic network" for representing the content of research/teaching fields. I showed that *such a network is a needed complement to traditional approaches since it is required by the following tasks which are crucial for education and research*:

- permitting students or researchers to *efficiently find and compare* all the objects (categories or statements) related to a given object and hence *more easily understand, memorize or evaluate* this object; for example, starting from a given task, finding/seeing all its subtasks organized into a subtask hierarchy, with the arguments/objections for the use of a task organized into an argumentation hierarchy, with the source of each task/argument/objection also displayed; this example illustrates how a semantic network permits to structure information scattered in many documents or many pages of a document;
- permitting lecturers, researchers and students to (i) add new objects (e.g., an objection to an argument) to an information repository *without introducing redundancies or lessening its organization*, and hence without having to create a new document, (ii) annotate (e.g., ask for clarification) on *precise objects* rather than big chunks of data, or (iii) make *precise votes on precise objects* (e.g., on their originality or "usefulness");
- permitting the managers of a repository (or the creators of a course) to make *recommendations* via annotations (like any other users) instead of making *rather arbitrary selections* on its content. Thus, each user *can* exploit *all* the fine-grained contributions from all users for (i) making her own *evaluation of contributions or contributors* (students, lecturers, researchers, etc.) or (ii) *filtering out what she (the user) does not want to see* when searching the repository or selecting a subset of it for an application or a course. Hence, for example, a lecturer may prepare a particular course via a query selecting all the contributions that have a minimal degree of usefulness, that have arguments without objections associated to them, and that have been authored by people having a PhD related to the domain of the course.

Section II summarizes the reasons why current approaches do not permit the creation of such a network, and introduces the approach I designed to permit the collaborative creation of such a network. In Open Learning and Distance Education, providing the students with unambiguous, well organized and not overly restricted learning materials is particularly important, if only because obtaining or providing feedback is more difficult and takes much more time than in face-to-face courses. Even a *course related newsgroup* (where problems or information related to a particular course are discussed) would be *advantageously replaced by a cooperatively-built semantic network*. Examples of rationales for this claim are: (i) a separate document (or the course itself) would not have to be updated by the lecturer for answering requests-for-clarifications from the students or for keeping an up-to-date easy-to-search repository of the important information with associated rationales (announcements, problem resolutions, etc.), (ii) recurring discussions - or repetitions of previously stated information - would be avoided, and (iii) since the statements would have to be more much precise and argued, in case of conflicts the resulting "structured discussion" would permit the involved parties or external parties to better compare and evaluate the respective positions than when they are scattered in the messages of a newsgroup.

Although all this is true from a theoretical perspective, and thereby provides interesting lights on several topics of this journal issues (e.g., "E-learning and pedagogical challenges", "Impact of e-learning on social change", "Future trends",

"Methods of combining traditional learning and e-learning" and "Web-based learning"), implementing this approach raises some problems. The techniques I created to solve these problems are not yet sufficient for genuine scalability, especially from a social viewpoint (usability and adoption by a great number of lecturers and researchers). This is summarized in Section II. Furthermore, not all these techniques are *fully* implemented in my knowledge base (KB) server WebKB-2 (Martin, 2001; Martin et al., 2005) (a "KB server" permits Web users to update one or several shared knowledge bases). The point of this article is *not to discuss* these techniques and their theoretical advantages (I did this in (Martin, 2001; Martin & Eboueya, 2008); their *summaries* in this article are original albeit concentrating many ideas in few sentences). *Instead*, the point is to present an experiment of the use of these techniques as a complement to traditional learning materials for distance and face-to-face teaching purposes. Section III presents extracts of semantic networks that I created for three courses and that were complemented by students. Section IV gives an evaluation of this experiment by some students which shows that much work remains to be done for the approach to be viable but that hope is permitted.

## 2. Background

### 2.1. Insufficiency of Other Approaches

Approaches based on the *indexation of resources* are not scalable. My first argument about this point in Martin & Eboueya (2008) is: "The more statements a resource contains, and the more resources there are, the more these resources contain similar and/or complementary pieces of information, and hence the less the meta-data for each resource can be useful. Indeed, queries on such meta-data return lists of resources that are partially redundant or complementary with each other and that need to be manually searched, compared or aggregated by each user". Approaches based on *fully formal resources* are also insufficient since semi-formal or informal objects (conceptual categories or statements) are unavoidable at certain levels of generality for organizing and presenting a KB, and for *supporting* an incremental refinement of its content. In the general case, approaches based on *mostly informal resources* do not permit to manually represent or automatically extract and normalize the meaning of the informally described objects and relationships, and hence do not permit to exploit them for information retrieval/comparison purposes. Similarly, ontology matching techniques (Euzenat & Shvaiko, 2007) are intrinsically limited by the lack of information contained in the source KBs (even knowledge engineers often cannot second-guess the knowledge providers and establish precise semantic relations between objects from different KBs). Hence, usable approaches based on *informal resources that were mostly independently created* formal resources are not scalable either.

The insufficiency of these approaches explains why many projects nowadays try to allow the cooperative creation of large KBs (like Wikipedia but much more formal and organized), e.g., Ontowiki (ontowiki.net) and Freebase (freebase.org). However, they do not provide protocols that genuinely support cooperation between people: only two KB servers - Co4 (Euzenat, 2001) (not available anymore) and WebKB-2 (usable at http://www.webkb.org/) - seem to have knowledge editing/voting/evaluation protocols that support *loss-less* knowledge integration (i.e., an integration that does not impose the users or managers of a KB to make a choice between inconsistent statements and hence loose knowledge; such choices can often only be made in the context of particular applications). Knowledge integration methodologies (e.g., Diligent, Dogma, HCome,

Methontology) or knowledge integration servers on the Web (e.g., Knowledge Zone) or in peer-to-peer networks (e.g., SomeWhere, CoAKTinG) impose choices during knowledge integration and hence are oriented toward the creation of applications rather than towards the cooperative creation of knowledge repositories.

## 2.2. Quick Overview of the Adopted Approach

In WebKB-2, every object (word, conceptual category, formal/informal statement, relation between objects) has one or several associated origins or believers (which are recorded objects too and hence can be used in statements and queries): 1) the user who created the object, 2) the source (e.g., a person, a language, a document) where the user read and hence *interpreted* the object (word or statement), and 3) other users which also believe in that object (statement). Lexical conflicts are avoided by prefixing category identifiers with the identifier of their creators (e.g., wn#bird refers to the most common concept proposed by WordNet for the word "bird"). For each new KB, WebKB-2 proposes a large general default ontology which is a loss-less integration of (i) many top-level ontologies and (ii) my conversion of WordNet into a genuine lexical ontology with intuitive identifiers (Martin, 2003). WebKB-2 also proposes various complementary notations (FCG, Formalized-English and FL) (Martin, 2002) that I derived from the Conceptual Graph Linear Form (CGLF) to further improve on what made its success: its readability, expressiveness and normalizing aspect (i.e., the fact that this notation helps people to represent statements in ways that ease the automatic finding of logical relations between them).

For redundancies or inconsistencies to be made explicit (and, from a logic-oriented viewpoint, removing semantic conflicts) and for keeping a minimal organization in the KB, before being added to the shared KB, each new object must be connected to at least one already existing object by a "*corrective*" relation (to state that the new object corrects an already existing object) or a "*generalization*" relation. (Between simple statements, a generalization relation corresponds to the logic implication. However, I also defined *extended generalization* relations not only to take into account constructs such as numerical quantifiers, sets or contexts but also to relate formal objects with informal objects, e.g., conceptual categories with words). Using *graph-matching* techniques, WebKB-2 can detect many partial/complete redundancies and inconsistencies between a new statement and those existing in the KB, and thus can ask the author of the new statement to refine it or add a corrective/generalization relation. For example, assuming that John has already entered in a semi-formal way that "all birds fly" and that John wants to enter that "most healthy French birds are able to fly", here is the semi-formal statement that Joe has to write if he uses the Formalized-English (FE) (Martin, 2002) notation currently usable in WebKB-2: `any bird is agent of a flight'(John) has for corrective_restriction `most healthy French birds are able to be agent of a flight' (Joe). In other words: Joe believes that his belief is a correction and restriction of John's belief. WebKB-2 also proposes a system to evaluate contributions and contributors based on votes and the way statements have been argued for or against (Martin et al., 2005). In the future, this evaluation system will be adaptable by each user, FE will be made more readable, and heuristics will be used for discovering semantic conflicts between statements even when they are informal.

This approach, along with the related induced use of precise formal categories (e.g., pm#Paris_in_1951 which specializes pm#Paris_between_1950_and_1960, which itself specializes pm#Paris) permits to put every imaginable belief into a same organized KB and avoid the problems related to integration s that are not loss-less (see Section II-A) and hence problems related to version control or truth-maintenance. When choices between conflicting beliefs have to be made, which is typically the case for applications but rarely for information retrieval, a selection of the knowledge to re-use can be made using queries and according to the characteristics of each application. For example, one application designer may select a consistent subset of a KB by selecting the most specialized and voted-for formal statements, and/or those that satisfy certain expressiveness constraints.

Finally, I also propose the use of replication mechanisms between competing or partially competing knowledge servers in such a way that it does not matter which server a user updates or queries first: the advantages of distribution and centralization are thus combined and there is only one "virtual" network (Martin et al., 2005).

### 2.3. Adoption of the Approach by Users and Tool Providers

There are various reasons why no other systems currently uses the approach adopted in WebKB-2 despite the advantages of this approach and the direct impact that its use would have on education. First, such a tool required much research and implementation work. Second, there currently exists a lot of informal legacy data but very little well-organized explicit knowledge. Third, the above described approach suffers from two problems common to all precision-oriented knowledge acquisition/retrieval approaches, i.e., approaches where the semantic network has to be (semi-)formal and displayed to the users: 1) people need to learn how to read such, and 2) entering knowledge representations requires much more intellectual rigor than writing informal sentences. The unwillingness of most people to learn new notations (e.g., musical/mathematical notations and programming languages) is well known. Furthermore, most people have not heard about knowledge representation languages nor about the usefulness of learning one.

Yet, I believe that my approach (combined with more traditional ones) has some future with researchers, teachers and students since (i) the need of using very small learning objects is now well recognized by the e-learning research community (Downes, 2001; Hodgins, 2006), (ii) the economy of time and resources brought by the use of truly re-usable learning objects will be understood by more and more e-learning/university teachers and administrators, (iii) more and more teachers are involved in e-learning, (iv) it is part of the roles of teachers and researchers to (re-)present knowledge in explicit and detailed ways, (v) my approach permits a better evaluation of the knowledge and analytic skill of the students than less precision-oriented approaches, and (vi) providing the semantic organization of the content of teaching materials (instead or in addition to these materials) help students find, compare and memorize the information scattered in these materials. As Section IV indicates, this last point was recognized by many of my students after they had learned how to read the semantic networks I prepared for them.

## 3. Presentation of the Created Semantic Networks

### 3.1. Content

During my e-learning fellowship (Martin, 2006a), I represented the content of three courses given at Griffith University (Australia). These courses were "Introduction to Multimedia Development" (*Multimedia*; 13 sets of slides), "Systems Analysis & Design" (*S.A.*; 437 slides) and "Workflow Management" (*WFM*; on-line course based on a book and supplementary materials). Most slides and most of their statements were represented into a semantic network of tasks, data structures, properties, definitions, etc. For example, 350 out of the 437 slides of the S.A. course were represented; only 350 because (i) the examples made via figures and tables were not represented, (ii) the redundancies were eliminated, and (ii) information solely related to tutorials and examinations was eliminated too. For the WFM course, at first the network only included the most important WFM concepts and relationships introduced in the book. This network was particularly helpful since in this book the descriptions of those important concepts and relationships were scattered amongst hundreds of sentences, and sometimes these descriptions were very general and fuzzy. Thus, without the network it was extremely difficult to remember and correlate all these descriptions.

For evaluating the students of these courses and the proposed approach, each student was asked to add at least twenty relations to the network. Then, I evaluated these additions (did they make sense? were they interesting? etc.). The WFM students had to do this exercise three times, as a replacement for a traditional learning journal.

For each of the three courses, a relatively small list of relation types happened to be necessary for representing the content of the slides and the relations between the important concepts. Most of these types were: subtype, instance, specialization, part (physical_part or subtask), technique, tool, definition, annotation, use, purpose, rationale, role, origin, example, advantage, disadvantage, argument, objection, requirement, agent, object, input, output, parameter, attribute, characteristic, support and url. (This list is ordered topically, not by frequency of occurrence.) This list is small compared to all the basic relations that can be found in top-level ontologies or that would potentially be needed if general natural language documents had to be represented.

### 3.2.   Presentation

The input files containing the initial knowledge representations for these courses are accessible at http://www.webkb.org/kb/it/. These input files were loaded into (i.e., executed by) WebKB-2 and hence their formal objects (conceptual categories or statements) became part of the unique global semantic network that can be queried, browsed and complemented by any Web user via WebKB-2 (http://www.webkb.org). The students were given the URL of WebKB-2 and the URLs of the input files for their courses.

Figure 1 shows an extract of an input file for the WFM course. Figure 2 shows the result of a very simple command. The figures 3 and 4 show extracts of input files for the Multimedia course. Within each input file the formal representations are included in sections and indented. This indentation most often reflects the specialization relations existing between the represented objects. The representations are enclosed within special tags (e.g., <script language="FL"> and </script>) to isolate them from the informal elements. HTML tags may be used within representations for presentation or hyper-linking purposes (Other HTML tags are ignored by WebKB-2; indexing a document element with representations is done in other ways). The figures 1, 3 and 4 show that knowledge from different topics can be represented, normalized and organized in very similar ways.

These input files use FL (Martin, 2002) because this notation was designed to be the most structured and concise possible formal notation that is as expressive as RDF+OWL-Full (Horrocks, 2003). (RDF+OWL is the knowledge representation language which has been recommended by the W3C and which has become the de-facto standard for the Semantic Web (Shadbolt & al., 2006) and hence the Semantic Learning Web too (Stutt & Motta, 2004; Devedzic, 2004)). FL is similar to N3 (Palmer, 2002) but has a more regular structure. (N3 is a readable notation often used in W3C documents to avoid using the XML-based notation for RDF+OWL.) FL is much more concise than other notations, especially graphic notations, and hence reduces the needs for scrolling or browsing. This permits people to see *many* relations between the formal objects, and hence better compare and understand these objects. This also eases the integration and exploitation of knowledge representations within textual documents or their connections with textual elements (Martin, 2006). An originality of the research work on is its focus on handy textual interfaces (notations, commands, and automatically generated textual interfaces). In the future, traditional solutions based on applets may also be used.

In the following examples (figures 1 to 4), no cardinalities are explicitly associated to the relations between the objects. Thus, each statement in these figures follow the generic schema "CONCEPT1 RELATION1: CONCEPT2 CONCEPT3, RELATION2: CONCEPT4, ...;". Such a statement should be read: "any CONCEPT1 may have for RELATION1 one or many CONCEPT2, and may have for RELATION1 one or many CONCEPT3, and may have for RELATION2 one or many CONCEPT4, ...". Some comments within the figures explain how the creators of each object are made explicit. As examples of additions made by students, please note the relations created by the student "s162557" in the figures 1, 2 and 4.

### 3.3. Ease of Creation

Thanks to FL and the large ontology of WebKB-2 (Martin et al., 2005) it was not too difficult to represent and categorize the important concepts and relationships contained in the source learning materials of the three courses. Furthermore, representing knowledge by extending a large shared ontology eases knowledge retrieval, re-use and understanding.

Although using a KB server such as WebKB-2 is unavoidable to allow the representation, querying and cooperative updating of a large semantic network, I found that a structured document editor (SDE; for example Amaya - the W3C Web browser - or any other XML editor) would have been a useful intermediary or complementary tool: (i) the manual creation of the representations would have been much easier if the source documents had been organized via a SDE instead of Word or Powerpoint, (ii) the manual exploitation of the input files would have been simpler with a SDE since for example some sections could have been temporarily hidden, and (iii) despite its semantic unawareness and use of predefined document schemas, a SDE could guide beginners in their creation of files and representations similar to the FL representations illustrated in the figures 1 to 4.

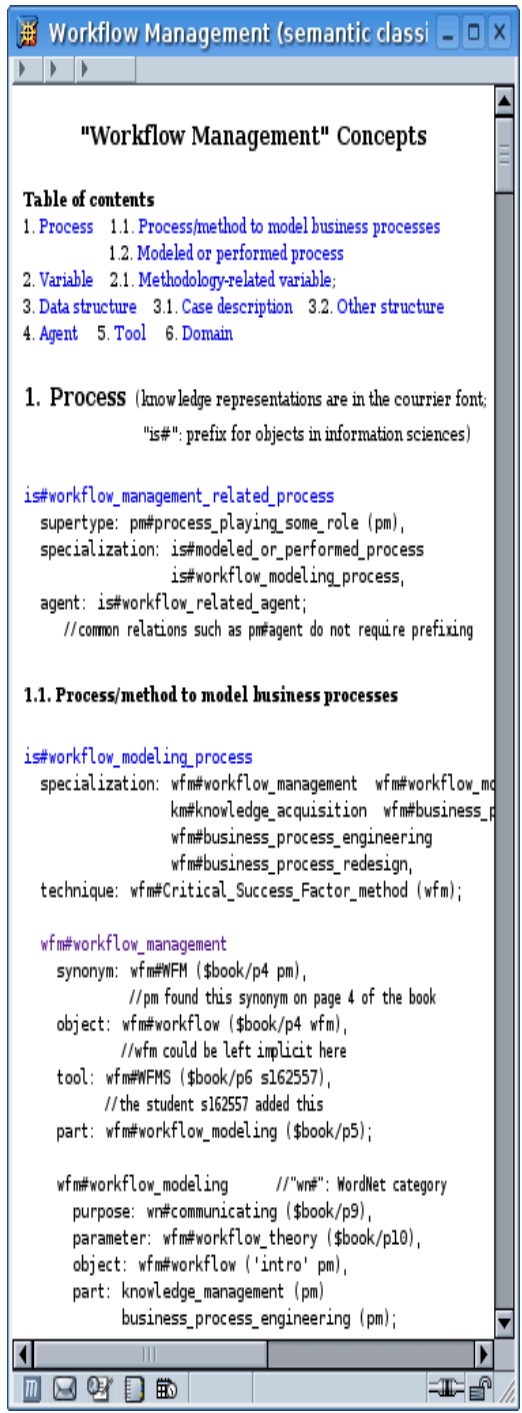

**Figure 1.** Extract from a file representing statements from a book in Workflow Management (here referred to by the variable $book)

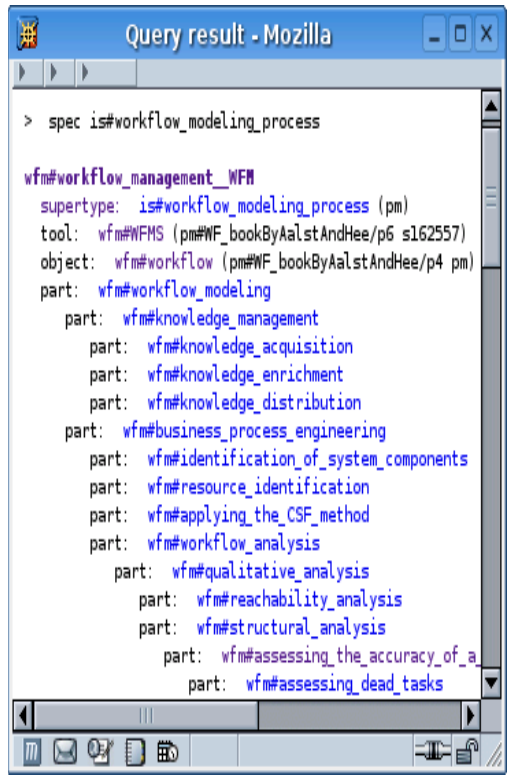

**Figure 2.** Command to display the specializations of a type, followed by its first result: wfm#workflow_management (here, this type is displayed along with some of its related objects using an informal format looking like FL)

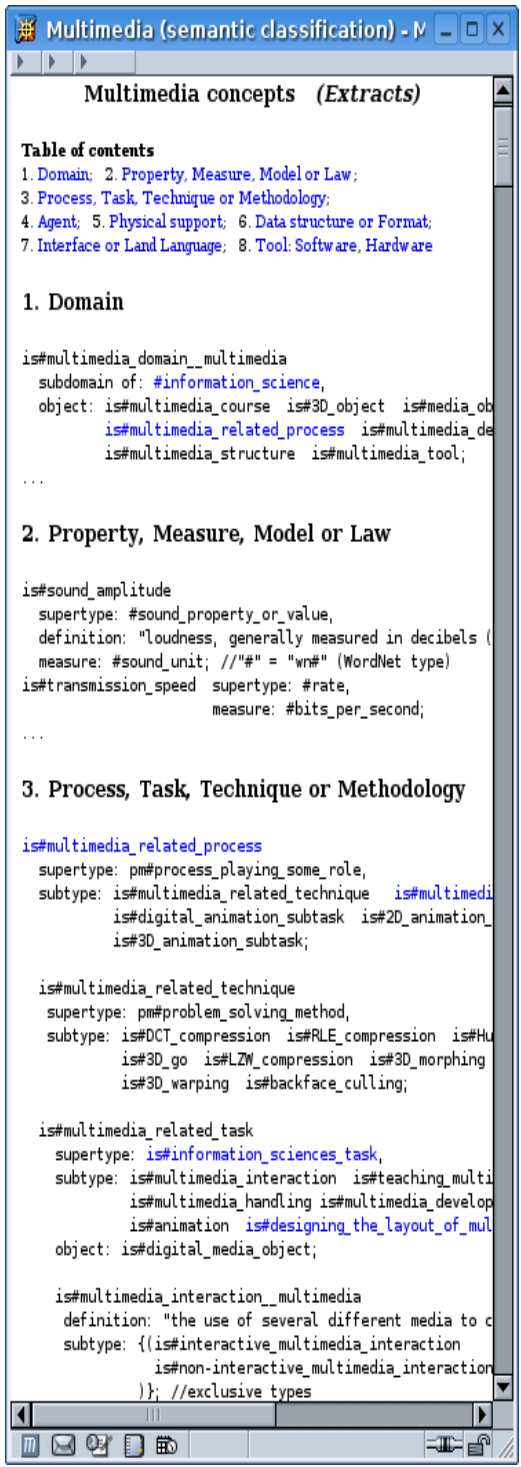

**Figure 3.  Organization of guidelines about the creation of Web pages**

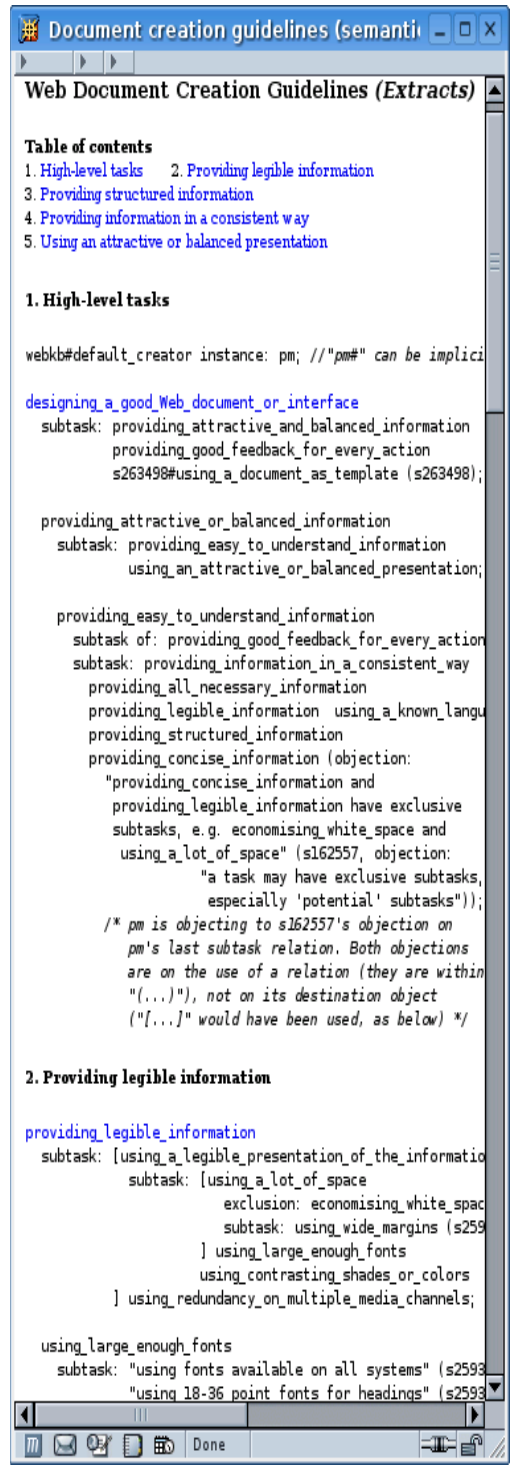

**Figure 4.  Extract from a file representing almost all statements of a course on multimedia**   (please read all lines carefully)

### 3.4.    Representations of Argumentation Structures

The students of the Multimedia course were also given a semantic network organizing "guidelines about how to create Web documents" and were asked to complement it based on guidelines that they would find on the Web (see Figure 4). The interest of this exercise came from the partial redundancy and unorganized nature of the many statements or Web pages that can be found on the Web about this subject.

The guidelines were first represented via informal statements including the word "should" and connected by formal argumentation relations. This proved not scalable: those statements could not be organized by specialization relations or any other relation permitting to create a hierarchy with a unique (onto)logical place[1] for each statement. A "unique place" (and hence the organization of statements into hierarchies) is a necessary condition for a user to know where to find or insert objects - as in a good decision tree - and hence for the hierarchy to remain useful as it grows.

An initial test with a small number of students quickly proved this last point: the initial argumentation hierarchy was not completed in a meaningful way by the students (the relations they added generally did not make sense) and the final hierarchy was hardly structured (it was nearly a flat list). The solution was quickly found: whenever semantically possible, the guidelines were represented again as a set of tasks and organized by subtask relations. This provided an intuitive structure for the students to complete and permitted them to strongly reduce their use of argumentation relations.

As alluded to in Section II *and illustrated in Figure 4*, my approach includes those of argumentation-based collaboration tools (e.g., (Uren et al., 2006)) but also allows (i) more expressiveness when required (e.g., *relations on relations*), (ii) the recording of votes and object creators for filtering or evaluation purposes, and (iii) a more normalized representation of knowledge (Martin & Eboueya, 2008).

### 3.5.    Use of a Wiki

I used WebKB-2 to create the semantic networks and the students used it for browsing and occasionally for additions. However, the students of the WFM and Multimedia courses had to be allowed the use of a classic wiki in addition to WebKB-2 for entering new concepts or statements. Indeed, no time was allowed for training the students to use FL in a syntactically correct way nor for giving them any real introduction to "knowledge representation" (the students were only shown how to read the representations and to avoid some ontological non-senses). Whenever a wiki was used instead of WebKB-2, most of the additions made by most contained lexical errors (e.g., typos or badly formed identifiers), syntactic problems (this is understandable), ontological problems (redundancies, inconsistencies, meaning-less relationships) and indentation problems. In Martin (2006b), I give a detailed list of errors made by the students of the WFM course in their first "semantically structured learning journals".

---

[1]    A "unique place" in an object hierarchy (e.g., a subtask hierarchy) does not imply that the hierarchy has to be a tree (e.g., a task can be subtask of many other tasks), it simply means that inserting a particular object is not "arbitrary". For example, a topic can be inserted at different places by different persons, depending on their goals, because the relation "sub-topic" does not provide enough conceptual constraints. Relations from the part of and specializationOf families have stronger constraints and hence different users are much more likely to insert (or look for) an object at the same place when they have a similar understanding of this object and of related existing objects in the hierarchy. As explained in SectionII, conceptual constraints and cooperation protocols can be used to lead the users to incrementally precise knowledge for correcting and avoiding incompatible understandings.

## 4. Quick evaluation of the Approach by Students

### 4.1. Presentation of the Survey Summary

Although conducting a systematic survey was part of the e-learning fellowship granted to me, for administrative reasons only the undergraduate students of the Multimedia course could finally be *invited* to fill a survey on their use of a semantic network. The summary of the anonymous survey that I conducted is shown by Table 1. While reading this table, it should be remembered that the students were poorly motivated to work on this experiment: they had no previous knowledge of "knowledge modelling" issues, most of them were only 18 years old, and they could not be marked nor given extra points for their participation in the survey. The 63 registered students of the Multimedia course were invited to study the proposed semantic network for their course, complete it (as part of their homework) and fill the survey. However, only 18 students actually did so, essentially those that I could ask to do so during a tutorial. Most of them only had a cursory a look at the proposed semantic network before adding a few relations and completing the survey. No more than 3 students submitted their homework before the deadline, that is, before the tutorial. The students' lack of participation should not be interpreted to reach any conclusion about the approach, it only shows that a systematic survey was not possible in my university (or, more precisely, its "School of ICT" department) given its administrative/marketing/financial constraints.

Since only 18 students of the Multimedia course filled their surveys, each of these students is indexed with an alphabetic letter in Table 1. Uppercase letters are used for the students who claimed to have spent at least half an hour on the semantic network. Lowercase letters are used for the other students. Some lowercase letters are prefixed by an underscore to highlight the fact that the students represented by those letters spent very little time on the representations (e.g., "0 minute" according to c and _i). At least _m and _o considered their answers irrelevant given how little they understood the representations and how little time they spent trying to understand them. The students referred to via lowercase letters (with or without underscore before) apparently also spent little time in completing their surveys too; for example, many questions are left unanswered and a "yes" answer has been given by _c, d and f to questions such as "which characteristics did you find (most) helpful?". It is interesting, encouraging and a bit surprising that apparently *only very few students had a negative or neutral attitude toward the proposed approach*, and that all of *these* students are referred to via a lowercase letter and often prefixed by an underscore. A question mark is used in the following figures when the answer was "I don't know" or "not sure".

The reader might be surprised by the format of this survey summary. As with FL, the advantages of this format come from its conciseness which permits to see and compare a lot of information. The adopted indentation/layout and indices provide a relatively tabular presentation (i.e., it is not difficult to find and compare all cells from a same row or column). A classic table would have been impractical to display, and a list of tables (or worse, individual surveys) would not have permitted easy comparisons. For the same reasons, graphics showing statistics would *not* have been advantageous to use. Showing statistics or making a statistical analysis would also have been *pointless* given the low number of interviewed students and the relatively large number of answers per student. Directly reading and comparing the answers in Table 1 is here *sufficient* to draw lessons from this experiment. Interestingly, given the students' answers, it seems that interviewing many more students would *not* have led to many more insights or deeper insights.

**Table 1: Evaluation of the given complementary materials by 18 students**

Age: A:24, B:18, _c:18, d:21, e:27, f:18, G:18, H:18, _i:17, J:18, k:18, l:23
   _m:22, N:21, _o:18, p:31, q:41, R:18

*Sex*: A:M, B:M, _c:M, d:M, e:M, f:M, G:M, H:M, _i:M, J:M, k:M, l:M,
   _m:F, N:F, _o:M, p:M, q:F, R:M

*Background (default: Australia):* B:Croatia, _c:Turkey, d:Singapore,
                    _m:USA, N:NZ, _o:Turkey, q:Canada

*If you have a job, how many hours do you work per week?*
 A:25, B:25, _c:15, d:20, e:15, f:25, G:25+, H:18, _i:14, J:25, k:15+, l:25,
 _m:20, N:10,   _o:15, p:12, q:10, R:

*How long (in minutes) have you spent on the complementary material?*
 A:180, B:, _c:0, d:20, e:15, f:, G:60?, H:180, _i:0, J:30, k:5, l:,
 _m:10, N:180,   _o:2, p:20, q:120, R:60-120

*How many times did you get back to this above cited document?*
 A:5-6, B:3, _c:0, d:0, e:1, f:, G:1, H:2, _i:0, J:0, k:1, l:,
 _m:0, N:3,   _o:0, p:, q:2, R:0

Did you feel that the complementary material was useful?

*1) did it help you understand relations between the objects?*
   A:yes, B:yes, _c:, d:yes, e:no, f:yes, G:no, H:yes, _i:no, J:yes, k:yes, l:yes,
   _m:no, N:yes,   _o:no, p:no, q:yes, R:yes

*2) did it help you better memorize the relations between the objects?*
   A:yes, B:no, _c:, d:yes, e:no, f:yes, G:no, H:yes, _i:no, J:yes, k:yes, l:yes,
   _m:no, N:yes,   _o:no, p:no, q:yes, R:yes

*3) did it help you find some relations in a quicker way?*
   A:yes, B:yes, _c:, d:yes, e:no, f:yes, G:?, H:yes, _i:no, J:no, k:yes, l:yes,
   _m:no, N:yes,   _o:yes, p:?, q:yes, R:yes

*4) did you find querying an object (or navigating from an object) helpful?*
   A:yes, B:yes, _c:, d:yes, e:no, f:yes, G:no, H:yes, _i:, J:, k:, l:yes,
   _m:no, N:yes,   _o:yes, p:, q:?, R:yes

*5) which characteristics did you find (most) helpful?*
   A:"relationships and examples of use", B:"the many comments",
   _c:, d:yes, e:none, f:yes, G:?,
   H:"easy navigation and good descriptions of scripts and objects",
   _i:, J:, k:"query boxes, links", l:notes, _m:none, N:yes,
   _o:"hierarchy system", p:, q:?, R:"simple lists"

*6) which characteristics did you find (most) unhelpful?*
   A:"confusing layout", B:"lack of spacing", _c:yes, d:, e:all, f:yes, G:?,
   H:none, _i:, J:, k:"sometimes too many relations", l:"can be confusing",
   _m:, N:none,   _o:"plain website", p:, q:"editing freely in the wiki",
   R:"information was a little scattered"

**Table 1 (con't): Evaluation of the given complementary materials by 18 students**

*7) do you have other comments or suggestions?*
A:"more verbose instructions and explanations for use",
B:"more color and comments would have helped",
_c:, d:, e:, f:, G:, H:, _i:, J:, k:"I also did pretty good without this", l:,
_m:"my answers are rather irrelevant since I did not understand the
    given document", N:"i know what they are now",
_o:"i haven't really used the site and i'm still passing",
p:"It helped me little. It was like trying to learn a new language",
q:"most of the notations were intuitive or well known",
R:"it is a very good tool for future use"

*8) do you wish you had this document sooner?*
A:yes, B:"no, It did come in handy", _c:, d:yes, e:, f:?, G:yes, H:yes,
_i:no, J:yes, k:yes, l:"no i found it useful to have before the test",
_m:yes, N:yes,   _o:no, p:, q:yes, R:no

Given your current understanding of the approach, imagine a better
tool interface for it. With this interface in mind, please answer the
following questions.

*1) would the approach (as a complement to the course) help you better
   understand the relations between the objects presented in this course?*
A:yes, B:yes, _c:?, d:yes, e:no?, f:yes, G:?, H:yes, _i:no, J:?, k:, l:yes,
_m:yes, N:yes,   _o:no, p:, q:, R:yes

*2) would the approach help you better memorize the relations?*
A:yes, B:no, _c:?, d:yes, e:no?, f:yes, G:?, H:yes, _i:no, J:yes, k:, l:yes,
_m:no?, N:yes,   _o:no, p:, q:, R:yes

*3) would the approach help you find a certain relation in a quicker way?*
A:yes, B:yes, _c:?, d:yes, e:no?, f:yes, G:?, H:yes, _i:no, J:yes, k:, l:yes,
_m:no, N:yes,   _o:no, p:, q:, R:yes

*4) would you find querying/navigating from an object helpful?*
A:yes, B:yes, _c:, d:yes, e:no, f:yes, G:yes, H:, _i:, J:yes, k:, l:yes,
_m:yes, N:,   _o:no, p:, q:, R:yes

*5) which characteristics would you find helpful (or most helpful)?*
A:"query function, annotation and examples", B:"see above", _c:, d:yes,
e:none, f:information, G:?, H:"easy navigation", _i:, J:, k:,
l:"the fact that it is well ordered with notes", _m:, N:,   _o:, p:, q:, R:

*6) which characteristics would you find unhelpful (or most unhelpful)?*
A:"layout ", B:"see above", _c:, d:yes, e:, f:complexity, G:?, H:none,
_i:, J:, k:, l:, _m:, N:,   _o:, p:, q:, R:

*7) other comments or suggestions?*
A:no, B:no, _c:, d:, e:, f:no, G:, H:, _i:, J:, k:, l:,
_m:"I don't understand programming terminology easily
    so having it merely listed is no good for m", N:,
_o:"make it more attractive", p:, q:, R:

**4.2.     Analysis**

My main interpretation of this survey is the following: (i) all the students who did their homework (even if they did not submitted it before the tutorial) found the approach useful (even if they did not like the syntax/layout), and (ii) many of the students who did not do their homework also understood the general advantages of the approach even if they did not *fully* understand the approach itself. The student referred to as "R" in the above summary made the following comment: "The time I spent researching the Web for design principles proved to be very useful to me. I now have a clearer knowledge of the principles after sorting them into the required format. The process was time consuming but did help me to remember and organize the concepts. This could be used in other subjects to help study for exams or just to process information".

In this experiment, no significant difference could be noted between the average mark of all students and the average mark of the students who were given access to the semantic network only after the mid-term exam.

WebKB-2 purposely uses textual interfaces rather than graphical interfaces (see Section III-B). This experiment tested  the acceptation of this approach with undergraduate students.  As expected, the sole use of FL for displaying the semantic network was a problem, although curiously one student (the one referred to as "q" above) thought that "most of the notations were intuitive or well known". Using controlled languages (e.g., Formalized-English) would not be a solution since these languages cannot display information in a sufficiently structured way. Although the use of FL with a good indentation leads to a structured display, this structure appears to be not explicit enough for beginners: understanding the structure and scope of the described relations was the students' main problem. Although less concise (i.e., more space-consuming) than FL, an interface based on structured elements (e.g., XML elements or embedded HTML tables) with specific background colours - and menus associated to each element - seems necessary for beginners to immediately understand the structure and scope of the described relations - and complement them more easily. However, precise knowledge representations *necessarily* include elements such as cardinalities, quantifiers, sets or contexts, and therefore still require the use of a *special notation* to express them as well as their scopes (indeed, structured elements are of no help for displaying such additional intertwined scopes).

**5.   Conclusions**

This article presented a scalable approach for precision–oriented knowledge sharing, its interest for open learning and distance education, and its application for this purpose. The student evaluation of this application seem to confirm our analysis in Martin & Eboueya (2008) (summarized in Section II of this article) indicating that the greatest obstacles against the use and cooperative building of a precise Semantic Learning Web are more social than technical (Section II-B lists technical solutions, Section II-C lists social obstacles). However, despite the low number of interviewed students, the survey clearly shows that even un-motivated, immature and knowledge modelling unaware students grasp the interest of the approach for complementing traditional learning materials.

WebKB-2 has various input-output formats and many presentation options but the students evaluation shows that    an additional format exploiting structured document elements seems necessary. The full implementation of the interfaces and mechanisms of WebKB-2 permitting its users to cross-evaluate each other's statements also need to be completed urgently. Finally, it appears essential to complement the cooperation protocols (Martin et al., 2005, 2008) with much stronger mechanisms to detect inputs that are either semantically incorrect or potentially redundant/contradictory with already existing statements. On the other hand, enhancing the search and browsing methods did not prove urgent and no user model seemed required: displaying large amounts of well structured information as query/navigation results appeared sufficient to let the users quickly find the information they want.

The temporary use of a wiki in addition to WebKB-2 was interesting because it confirmed how inadequate wikis are for (i) letting people collaboratively build structured knowledge, and (ii) evaluating their contributions. Indeed, the ease-of-use of wikis does not compensate for their lack of semantic structure, semantic checking and cooperation protocols. Current semantic wikis are only timid advances toward the support of semantic structures/checking. Apart from OntoWiki (Auer et al., 2006) which includes the features of a frame-based system, most semantic wikis often offer very little support for fine-grained systematic knowledge modelling. For example, within a page, Semantic MediaWiki (Krötzsch et al., 2007) only allows to set semantic relations from/to the object represented by the page, and only in a rather hidden way within an unstructured text. The reasons why semantic wikis and most KB servers should have – but do not yet have - editing/voting/evaluation protocols supporting loss-less knowledge integration have been given in Section II.

Although the student evaluation is interesting at various levels, I mainly interpret it as a confirmation of the urgency of implementing more features. Unlike data management tools, KB management tools cannot be built by combining small independent tools, they must be full-featured to be adopted. Limiting their number of features to reduce their complexity is not a winning strategy (Shipman & Marshall, 1999), however tempting and popular it may be. This is especially true to achieve the constraint of "scalability", that is, to reduce future extension problems and keep guiding users as the KB grows.